\documentclass[12pt]{article}
\newcommand{\newc}{\newcommand}
\newc{\gsim}{\lower.7ex\hbox{$\;\stackrel{\textstyle>}{\sim}\;$}}
\newc{\lsim}{\lower.7ex\hbox{$\;\stackrel{\textstyle<}{\sim}\;$}}
\newc{\gev}{\,{\rm GeV}}
\newc{\mev}{\,{\rm MeV}}
\newc{\ev}{\,{\rm eV}}
\newc{\kev}{\,{\rm keV}}
\newc{\tev}{\,{\rm TeV}}

\newc{\mz}{M_Z}
\newc{\mpl}{M_*}
\newc{\mw}{m_{\rm weak}}
\def\beq{\begin{equation}}
\def\eeq{\end{equation}}
\def\bea{\begin{eqnarray}}
\def\eea{\end{eqnarray}}
\newc{\ie}{{\it i.e.}}          \newc{\etal}{{\it et al.}}
\newc{\eg}{{\it e.g.}}          \newc{\etc}{{\it etc.}}
\newc{\cf}{{\it c.f.}}
\def\inv{^{\raise.15ex\hbox{${\scriptscriptstyle -}$}\kern-.05em 1}}
\def\lbar{{\lower.35ex\hbox{$\mathchar'26$}\mkern-10mu\lambda}} 

\let\<=\langle
\let\>=\rangle

\let\+=\uparrow
\renewcommand{\epsilon}{\varepsilon}
\renewcommand{\phi}{\varphi}
\begin{document}
\thispagestyle{empty}
\vspace*{.5cm}
\noindent
\hspace*{\fill}{\large CERN-TH/2001-378}\\
\hspace*{\fill}{\large MIT-CTP \# 3236}
\vspace*{2.0cm}
\begin{center}
{\Large\bf Depilating Global Charge From Thermal Black Holes}
\\[2.5cm]
{\large John March-Russell$^1$ and Frank Wilczek$^2$}\\[.5cm]
{\it $^1$Theory Division, CERN, CH-1211 Geneva 23, Switzerland}
\\[.2cm]
{\it $^2$Center for Theoretical Physics, MIT, Cambridge MA 02139-4307, USA}
\\[1.1cm]
{\bf Abstract}\end{center}
\noindent
At a formal level, there appears to be no difficulty involved in
introducing a chemical potential for a globally conserved quantum
number into the partition function for space-time including a black
hole.  Were this possible, however, it would provide a form of black
hole hair, and contradict the idea that global quantum numbers are
violated in black hole evaporation.  We demonstrate dynamical
mechanisms that negate the formal procedure, both for topological charge
(Skyrmions) and  complex scalar-field charge.  Skyrmions collapse to
the horizon; scalar-field charge fluctuates uncontrollably.     
\newpage
\setcounter{page}{1}

\section{Introduction}
The classic no-hair theorems of black hole physics~\cite{nohair} are commonly
interpreted as implying that (non-gauge) global conservation
laws are inevitably violated in the process of black hole
evaporation~\cite{Hawking},
even in the absence of any explicit microscopic mechanism
for such violation~\cite{TTWZ,KW,Coleman,KMR,Holman,Wheeler}. 
If black hole evaporation, and more generally quantum gravity does
violate global conservation laws then this has many implications for
the use of global symmetries in fundamental physics, including
baryon and lepton-number violation, axion physics, the use of very
light scalars in cosmology, and models of low-scale quantum
gravity~\cite{TTWZ,KW,KMR,Holman,ADMR}.
But since the no-hair theorems are essentially classical,
and they are derived strictly only for stationary geometries,
which contain unresolved singularities, perhaps some doubt remains
possible.  

An interesting alternative perspective is afforded by passing to
imaginary time, and considering Euclidean black holes.  The
Euclidean Schwarzschild metric
\beq
ds^2 ~=~ \left(1-\frac{2M}{r}\right) d\tau^2 +
\left(1-\frac{2M}{r}\right)^{-1} dr^2 +
r^2 d\Omega^2
\eeq
appears to have a singularity at $r=2M$, but upon substituting
$v= \sqrt{8M} \sqrt{r-2M}$ one finds that near this limit
\beq
ds^2 ~\rightarrow~ \frac{1}{16M^2} v^2 d\tau^2 + dv^2  + (2M)^2 d\Omega^2,
\eeq
and by enforcing the periodicity $\tau/4M = \tau/4M + 2\pi$
we match non-singular polar coordinates.  Two consequences of this
construction are that the region behind the horizon, $r< 2M$, does
not appear; and that we are describing an appropriate, non-singular
background for quantum field theory at temperature
$T= (8\pi M)^{-1}$~\cite{GH}.   

Indeed, if one considers formally the thermal partition function 
\begin{equation}
Z~=~ \int {\cal D} g {\cal D} \phi e^{-S_\beta} 
\end{equation} 
including gravity and generic matter fields $\phi$, and
integrating over fields periodic in the imaginary
time period $\beta$, the Euclidean Schwarzschild solution appears
as a stationary point.   

Superficially it
appears innocuous to add a chemical potential term, for any
microscopically conserved
global charge, into this definition.  But if by doing so we
found a meaningful partition function, depending on the chemical
potential, we would be able to define a conserved charge
on the black hole, and contravene the above-mentioned conventional
wisdom~\cite{Coleman}.  In the remainder of this note we shall
consider two different
sorts of global charge, topological and ordinary, and demonstrate
that in both cases something goes wrong with this attempt --
quite different things in the two cases.

\section{Skyrmion Collapse}

Let us briefly recall the construction of Skyrmions\cite{Skyrme}.
Consider the non-linear $\sigma$ model defined by the
four-component scalar field $\phi^a$, $a=0,\dots , 3$ with $\phi^2 = 1$;
this defines a target space $S^3$.    At spatial infinity
the field approaches a uniform value, say $\phi^0 = 1$, defining the
normal vacuum.   Field configurations approaching a
constant at spatial infinity define maps $S^3\rightarrow S^3$.  
There is a topological current density
\begin{equation}
j^\mu ~=~ \frac{1}{3\pi}\epsilon^{\mu \alpha \beta \gamma}
\epsilon_{dabc} \phi^d \partial_\alpha \phi^a \partial_\beta
\phi^b\partial_\gamma \phi^c
\end{equation} 
which is identically conserved.   The charge obtained by integrating the
zero component of this current over space is the
degree of the associated mapping.  The static Skyrme textures 
\begin{eqnarray}
\phi^0 (\vec r ) &= &\cos f(r), \cr
\vec \phi (\vec r)& = &\sin f(r)~\hat r ,
\end{eqnarray}
define a class of symmetrical mappings with 
\begin{equation}
j^0 (r) ~=~ \frac{2}{\pi} f^\prime (\sin f )^2  ~=~ \partial_r
\frac{1}{\pi} \left( f - \frac{1}{2} \sin (2f) \right). 
\end{equation}
Regularity at the origin implies $f(0)= n\pi$.  If $f(r) \rightarrow 0$
at spatial infinity, arriving at the normal vacuum, then the charge is $-n$. 

Now let us consider the energetics, first with reference to flat space. 
If we use the standard non-linear $\sigma$ model Lagrangian 
${ \cal L} = \sqrt g \frac{1}{2} g^{\mu\nu}\partial_\mu \phi^a
\partial_\nu \phi^a$ then the energy $E_\lambda$ of a
re-scaled Skyrme texture
$f_\lambda (r) = f(\lambda r)$ transforms as $E(\lambda )= \lambda E(1)$. 
Thus we have charged configurations with
arbitrarily small energy, using singular maps whose structure is
concentrated near the origin (Derrick's theorem).    The
standard remedy is to supplement the Lagrangian with the
higher-derivative Skyrme term
\begin{equation}
{\cal L}_{\rm Skyrme} \propto - \sqrt g g^{\mu \rho}
g^{\nu \sigma} F^{ab}_{\mu \nu} F^{ab}_{\rho \sigma}
\end{equation}
with $F^{ab}_{\mu \nu} \equiv \partial_\mu \phi^a \partial_\nu \phi^b -
\partial_\nu \phi^a \partial_\mu \phi^b$.  A
special property of this term is that it contains no higher than
two powers of the time derivative, so that the action including
it  continues to define a normal dynamical system.   The energy
associated with this term scales as 
$E_{\rm Skyrme} (\lambda ) = \lambda^{-1} E_{\rm Skyrme} (1)$,
oppositely to the minimal term.  Thus it is no longer
favorable for field configurations to collapse.   Indeed, in
flat space one finds a non-trivial function $f$ with $n=1$ that satisfies
the equation of motion and minimizes the energy.  

In the Euclidean Schwarzschild background, both the topology and
the energetics of the situation are altered.  The topology of
spatial sections (constant $\tau$) is $S^2 \times R^+$ (a sphere times
a half-line), and of the entire space-time is $S^2 \times
R^2$; in neither case is there a quantized degree defined.  Indeed, in
the Skyrme texture, the restriction $f(0) = n\pi$ is no
longer required by continuity nor (for physical purposes, decisive) by
demanding finite energy.   This is because the factor
multiplying the angular part of the metric does not degenerate at the
origin, which is effectively the horizon $r=2M$.   

In more detail, the energy near the horizon for a Skyrme texture
parametrized by $f(r) = h(v)$ behaves as 
\begin{equation}
E \sim \int_0 dv ~v~ \left( c_1 ( h^\prime)^2  + c_2 (\sin h)^2 +
c_3 (\sin h)^2 (h^\prime)^2  + c_4 (\sin h)^4 \right) , 
\end{equation}
with numerical constants $c_i$.  (Here and below we refer to the
energy conjugate to Schwarzschild time, i.e. the action per unit $\tau$.)
The first two terms arise from the minimal $\sigma$-model
Lagrangian, the second two from the Skyrme term, for radial and angular
derivatives respectively.  To analyze this, it is convenient to switch
to the tortoise coordinate $s \equiv \log 1/v$, so 
\begin{equation}
E \sim \int^\infty ds \left( c_1 ( h^\prime)^2 + c_2 e^{-2s} (\sin h)^2 +
c_3 (\sin h)^2 (h^\prime)^2 + c_4 e^{-2s}
(\sin h)^4\right).
\end{equation}
No divergence arises when $h$ approaches an arbitrary
constant value at $s\rightarrow \infty$, so requiring
finite energy does not quantize the charge \cite{LM}.  Much more is true.
Consider a configuration with $h$ varying uniformly from $0$ to
$\zeta$ over the interval $[a,b]$.  The energy behaves as 
\begin{equation}
E \leq c_1 \frac{\zeta^2}{b-a} + c_2 e^{-2a} +
c_3 \frac{\zeta^2}{b-a} + c_4 e^{-2a} . 
\end{equation}
Since this can be made arbitrarily small by taking $a$ and $b$ suitably
to infinity, the mass/charge ratio for Skyrmion
charge is minimized at zero, by collapsing charge toward the horizon.  
Equivalently, defining $u=r-2M$, and taking $h(u) \sim (\log u)^p$ then
gives a most singular contribution of the form 
\beq 
E\sim \int du {(\log u)^{2p-2} \over u} 
\eeq
which is convergent for $p<1/2$.  Thus a logarithmic divergence in the 
ansatz function $h(r)$ is can occur with finite action.  This implies that 
the mass/charge ratio for Skyrmion charge is minimized at zero.

By way of contrast, in flat space the behavior of the energy near the
origin is
\begin{equation}
E \sim \int_0 dr ~r^2~ \left( \kappa_1 (f^\prime)^2 +
\kappa_2 \frac{1}{r^2} (\sin f)^2 + \kappa_3  \frac{1}{r^2} 
(f^\prime)^2 (\sin f)^2 +
\kappa_4 \frac{1}{r^4} (\sin f)^4 \right). 
\end{equation}
Quantization of the charge and non-triviality of the mass/charge ratio are
implicit in this form.   Indeed, finiteness of the $\kappa_4$ term
requires $\sin f(0) =0$; and some simple
arguments using Schwarz's inequality with the $\kappa_3$ term for
small $r$ and the $\kappa_1$ term
for large $r$ allow us to bound the charge in terms of the energy.  It
seems worth recording that without the Skyrme term there would be no
physical quantization of charge, since the $\kappa_1$ and $\kappa_2$ terms
permit non-zero $\sin f(0)$ at finite energy.

The possible accumulation of charge with zero energy near the horizon
reminds us of the increasingly red-shifted 
image that remains, in real space-time, as a record of whatever has
fallen into the hole.    On the other hand, such charge leaves
no long-time residue at any finite distance from the horizon, and in this
sense it does not provide hair.

\section{Scalar Field -- Limiting Mass/Charge Ratio}

Now let us consider the conserved charge associated with the phase
symmetry of a complex scalar field $\phi$.   For simplicity
we specialize to factorized $s$-wave configurations $\phi(\vec r, \tau)
= \eta (v) e^{-i n \tau/4M}$, taking account of the
periodicity in imaginary time.  Near the horizon the charge behaves as
\begin{eqnarray}
Q&=& \int \sqrt g g^{\tau \tau} {\rm Im} (\phi^* \partial_\tau \phi ) \cr 
&=& (4\pi (2M)^2 ) \frac{n}{4M} \int_0  dv~ \frac{1}{v} (\eta (v) )^2 \cr
&=&  (4\pi (2M)^2 ) \frac{n}{4M} \int^\infty ds~ ( \eta (s) )^2.  
\end{eqnarray}
Near the horizon the energy ($\equiv {\rm action}/{\rm unit~}\tau$)
behaves as 
\begin{eqnarray}
E&=& \int \sqrt g  (g^{\tau \tau } \partial_\tau \phi^*
\partial_\tau \phi +  
g^{vv} \partial_v \phi^* \partial_v \phi + m^2 \phi^* \phi) \cr
&=&  (4\pi (2M)^2 )\int_0 dv~  \left( \frac{1}{v} \eta^2
\left(\frac{n}{4M}\right)^2+
v (\eta^\prime)^2 + v m^2 \eta^2\right) \cr
&=&  (4\pi (2M)^2 )\int^\infty ds ~  \left( \left(\frac{n}{4M}\right)^2\eta^2 +
(\eta^\prime )^2 + e^{-2s} m^2 \eta^2 \right).  
\end{eqnarray}
By taking $\eta \rightarrow {\rm const.} s^{-\frac {1}{2} + \epsilon}$
as $s\rightarrow \infty$, with small $\epsilon$, we can
enhance the first term in the integrand relative to the other two.  
In the limit, the energy and charge become proportional, in the form
\begin{equation}
\frac{E}{Q} = \frac{1}{4M},
\end{equation}
using $n=1$ (the most favorable).  
The peculiar appearance of this equation arises from the fact that we have
put the Planck mass $M_{\rm Pl} =1$.  Restoring
units, we have $E/Q = M_{\rm Pl}^2/4M$.  For large black holes, this
becomes small.  Indeed, the ratio is of order the Hawking
temperature.  Thus
$E/Q$ is equal to the electron mass for $M \sim 2\times 10^{17}$ gm; for
a solar-mass black hole, the ratio is $E/Q \approx
6\times 10^{-16} m_e$.    So for macroscopic holes it becomes
energetically favorable 
to hide quantum numbers in a singular manner
near the horizon, similar to what we found in the Skyrme model.  In any
case, the charge near the horizon undergoes significant
fluctuations due to the Hawking temperature.   Without further analysis,
however, it is not clear how this phenomenon addresses
our problem of principle for small holes and heavy charge quanta.

\section{Scalar Field -- Charge Veto}

Referring again to the energy (or action) expression, we see that there
is a qualitative difference between the behavior of
the $n=0$ and the $n=1$ (or higher) sectors near the horizon.   In
both cases the mass terms become negligible, and we are left
to compare actions of the form $\int^\infty ds \kappa_2 (\eta^\prime)^2$
versus  $\int^\infty ds  (\kappa_1 \eta^2 + \kappa_2 (\eta^\prime)^2 )$.
The first form permits asymptotics $\eta \rightarrow
s^{+\frac{1}{2} -\epsilon}$ while the second requires the much more
stringent condition $\eta \rightarrow s^{-\frac{1}{2}
-\epsilon}$, so we might expect that the class of configurations it
supports has relatively small measure.  To quantify this, consider
the functional determinants accompanying integration over these sectors,
concentrating on the contribution to the action from an
interval $[a, a+L]$ in $s$.    For simplicity, take $\kappa_1 = \kappa_2 =1$,
remove the constant mode, and assume periodic boundary conditions.  This
is only a reasonable approximation to the contribution for short-wavelength
modes, which decouple from other intervals.   The relevant determinants
involve the product of inverse square roots of the eigenvalues, so for
their ratio $r$ we have the infinite product
\begin{equation}
r = \prod^{\infty}_{1}
\frac {(\frac {2\pi n}{L})^2 }{1 + ({\frac {2\pi n}{L})^2} } .
\end{equation}
This evaluates to $\frac{L} {2\pi}(\sinh (\frac {L}{2\pi}))^{-1}$.  
The low eigenvalues are not to be taken seriously, as already mentioned.
But the exponential falloff in $L$ arises from the high
eigenvalues, and it indicates that the  $n \neq 0$ sectors have zero measure
relative to the $n=0$ sector.  
As we have seen in the previous section, the charge operator itself
contains a term of the same asymptotic form as 
the $\kappa_1$ contribution.   Thus a chemical potential can decrease
the net numerical value of
$\kappa_1$, or even reverse its sign.  Therefore we arrive at two
simple possibilities.  If the chemical potential is less than the critical
value $1/4M$, it has no effect.  If it is greater than this value, the
action becomes unbounded below.   Neither of these possibilities 
yields a clear realization of a hairy black hole. 
Projection on charge eigenstates goes through the intermediary
of partition functions with chemical potential~\cite{Coleman},
and so encounters
the same singularity.  We cannot, from this
analysis, entirely preclude the possibility of some more
delicate construction, somehow working with chemical potentials
infinitesimally near the critical value, 
but it would require some additional ideas.

\section{Conclusion}

Straightforward attempts find thermal hair in canonical realizations of
topological and non-topological global charges appear to go awry, 
for different and perhaps not entirely straightforward reasons.  In both
cases, special properties of the Euclidean black hole metric allow one
to store charge in a singular fashion near the horizon of a large hole,
with small cost in energy.   

Throughout our discussion we have treated the background geometry as fixed,
thus neglecting possibility of ``back-reaction''.   The crucial configurations
for our arguments, in the Skyrmion case, involved fields varying rapidly
near the horizon.   Although their integrated action is small, they induce
large local values of the energy-momentum tensor.   Thus there is no
guarantee that they will correspond to valid configurations of the full
quantum gravity theory.  Indeed, if they did we would appear to be in
danger of providing an infinite entropy for the black hole, since there
is at least one low-energy state for each value of the charge.  
(This is related to the phenomena discussed by 't~Hooft~\cite{tHooft}.)

Finally, let us venture a heuristic interpretation of the charge veto.  The
Schwarzschild temperature $(8\pi M)^{-1}$ is appropriate for static frames
far from the hole, but at finite distances the effective local temperature
is $\sqrt {g^{\tau\tau}}$ larger, which diverges near the horizon.   Thus
the chemical potential becomes negligible compared to the local temperature,
and loses its influence upon configurations concentrated near the horizon,
unless it triggers an instability.  

\section*{Acknowledgements}
JMR wishes to thank Arthur Hebecker for useful conversations.  This
work is supported in part by funds provided by the U.S. Department of
Energy (D.O.E.) under cooperative research agreement
\#DF-FC02-94ER40818.

\end{document}